\documentclass[twocolumn,showpacs,aps,pra,eqsecnum]{revtex4}
\usepackage{graphicx}
\usepackage{dcolumn}
\usepackage{bm}
\usepackage{amsfonts}
\usepackage{latexsym}
\usepackage{amssymb} 
\usepackage{verbatim}
\usepackage{amsthm}
\usepackage{amsmath}

\def\fun#1#2{\lower3.6pt\vbox{\baselineskip0pt\lineskip.9pt
  \ialign{$\mathsurround=0pt#1\hfil##\hfil$\crcr#2\crcr\sim\crcr}}}
\begin{document}


\title{
  Re-interpretations of an experiment on the back-action in 
  a weak measurement
}

\author{
  Kouji Nakamura${}^{1}$
  and
  Masataka Iinuma${}^{2}$
}
\email{kouji.nakamura@nao.ac.jp}
\affiliation{%
  ${}^{1}$TAMA project, Optical and Infrared Astronomy Division,  
  National Astronomical Observatory of Japan,\\
  Osawa 2-21-1, Mitaka, Tokyo 181-8588, Japan\\
  ${}^{2}$ Graduate school of Advanced Sciences of Matter,
  Hiroshima University,\\
  1-3-1 Kagamiyama, Higashi-Hiroshima
  739-8530, Japan
}%


\date{\today}

\begin{abstract}
  Interpretations of an experiment on the back-action in a weak
  measurement in [M. Iinuma et al., New J. Phys. {\bf 13}
  (2011), 033041] are revisited.
  We show two different but essentially equivalent
  interpretations for this experiment along the original
  scenario of weak measurements proposed by Aharonov, Albert,
  and Vaidman.
  To do this, we introduce the notion of extended weak values
  which is associated not only with the states of the system but
  also the state of the measuring device.
  We also evaluate fluctuations in this experiment and found
  that an optimal measurement strength exists 
  for a fixed polarization angle prepared as an initial state, at which
  fluctuations in measurement results vanish.
  The consistency of this evaluation with the experimental results 
  is discussed.
\end{abstract}

\maketitle

\section{Introduction}
\label{sec:Introduction}


Weak measurements are topical subjects in modern quantum mechanics.
Since the proposal of the weak measurement by Aharonov, Albert,
and Vaidman (AAV)~\cite{Y.Aharonov-D.Z.Albert-L.Vaidman-1988} in
1988, many theoretical and experimental researches have been carried out.
The idea of weak measurement has been used to resolve fundamental
issues in quantum mechanics such as Hardy's
paradox~\cite{Y.Aharonov-A.Botero-S.Pospescu-B.Reznik-J.Tollaksen-2001},
violation of the Leggett-Garg
inequality~\cite{N.S.Williams-A.N.Jordan-2008,Y.Suzuki-M.Iinuma-H.F.Hofmann-2012},  
and the error-disturbance
relation~\cite{A.P.Lund-H.M.Wiseman-2010} derived by
Ozawa~\cite{M.Ozawa-2003}. 
On the other hand, many
experiments~\cite{N.W.M.Ritchie-J.G.Story-R.G.Hulet-1991,G.J.Pryde-J.L.O'Brien-A.G.White-T.C.Ralph-H.M.Wiseman-2005,O.Hosten-P.Kwiat-2008,P.B.Dixon-D.J.Starling-A.N.Jordan-J.C.Howell-2009}
show that the weak measurement is also very useful for
high-precision measurements.


The original proposal by AAV is based on
von-Neumann's measurement theory~\cite{J.von-Neumann-1932} in
which the whole system involves a quantum system to be measured 
and a measuring device to measure it.
According to the AAV scenario, weak measurements have  
four steps: the first one is the preparation of the initial
state of the system (pre-selection) as well as the measuring
device; the second one is to induce an interaction weakly 
between the system and the measuring device; the third one is
the selection of the final state of the system (post-selection)
by the projection measurement; and the final one is the
measurement of the state of the measuring device.
Through these four steps, the experimentally obtained value in
the {\it linear-order} of the weak interaction between the
system and the measuring device is not an eigenvalue of the
operator $\hat{O}$ of the system but the weak value of
the operator $\hat{O}$:
\begin{eqnarray}
  \label{eq:weak-value-def}
  \langle \hat{O}\rangle_{w}
  =
  \frac{
    \langle\psi_{2}|\hat{O}|\psi_{1}\rangle
  }{
    \langle\psi_{2}|\psi_{1}\rangle
  }
  ,
\end{eqnarray}
where $|\psi_{1,2}\rangle$ are the pre- and post-selected states
of the system, respectively.


Along this AAV scenario of weak measurement, one of the authors
revealed some properties of weak measurements taking into
account of non-linear effects of the von-Neumann
interaction~\cite{K.Nakamura-A.Nishizawa-M.-K.Fujimoto-2012}.
In Ref.~\cite{K.Nakamura-A.Nishizawa-M.-K.Fujimoto-2012}, the
arguments are restricted to an operator $\hat{O}$ which satisfies 
the property
$\hat{O}^{2}=\hat{1}$ and the {\it all-order} evaluations with 
respect to the coupling constant in the von-Neumann interaction
$H=gp\hat{O}\delta(t-t_{0})$ have been carried out.
As the results, it is shown that the final results of weak
measurements are given by 
\begin{eqnarray}
  \frac{\langle q\rangle'}{g}
  &=&
  \frac{
    \mbox{Re}\langle\hat{O}\rangle_{w}
  }{
    1
    +
    \frac{1}{2}
    \left(
      1
      -
      \left|\langle\hat{O}\rangle_{w}\right|^{2}
    \right)
    \left(
      e^{-s} - 1
    \right)
  }
  ,
  \label{eq:kouchan-A2is1-52}
  \\
  g \langle p\rangle'
  &=&
  \frac{
    s e^{-s} \mbox{Im}\langle\hat{O}\rangle_{w}
  }{
    1
    +
    \frac{1}{2}
    \left(
      1
      -
      \left|\langle\hat{O}\rangle_{w}\right|^{2}
    \right)
    \left(
      e^{-s} - 1
    \right)
  }
  .
  \label{eq:kouchan-A2is1-53}
\end{eqnarray}
Here, $q$ and $p$ are the canonical variables for the measuring 
device ($[q,p]=i$, $\hbar=1$), the initial state of the
measuring device was assumed to be the zero mean-value Gaussian,
and $s$ is a parameter of the measurement strength defined by
$s:=2g^{2}\langle p^{2}\rangle$ with the initial momentum
variance $\langle p^{2}\rangle$ of the measuring device,
$\langle q\rangle'$ and $\langle p\rangle'$ are the expectation
value of the pointer variable and its conjugate momentum after
the post-selection.


On the other hand, the experimental group of one of the
authors~\cite{M.Iinuma-Y.Suzuki-G.Taguchi-Y.Kadoya-H.F.Hofmann-2011}
realized a weak measurement of a photon polarization close to
ideal and showed that the back-action effects in the weak
measurement have an important role in the regime where the weak
value becomes large.
We refer this reference as
ISTKH~\cite{M.Iinuma-Y.Suzuki-G.Taguchi-Y.Kadoya-H.F.Hofmann-2011}
in this paper.
In ISTKH, it is claimed that the weak value which obtained by
the experiment is different from Eq.~(\ref{eq:weak-value-def}).
The ``experimental weak value''
$\left(\langle\hat{O}\rangle_{w}\right)_{exp}$ in ISTKH is given
by
\begin{eqnarray}
  \label{eq:Pb1-Pb2_in_ISTKH2011-0}
  \left(\langle\hat{O}\rangle_{w}\right)_{exp}
  &=&
  \frac{
    \langle\hat{O}\rangle_{w}
  }{
    1
    +
    \eta \left(
      \left|\langle\hat{O}\rangle_{w}\right|^{2} - 1 
    \right)
  }
  ,
\end{eqnarray}
where $\eta$ is a parameter describing the back-action effect.
In ISTKH, the expression of
Eq.~(\ref{eq:Pb1-Pb2_in_ISTKH2011-0}) is derived from the
formalism of the positive operator valued measure (POVM) and
the combination of unitary transformations representing optical
components, but this derivation does not directly follow the
original scenario proposed by AAV.
In particular, the ISTKH experiment explicitly utilizes a
maximally entangled state of the system and the measuring device,
which is produced before the weak interaction between them.
In spite of the above difference, the results
Eq. (\ref{eq:kouchan-A2is1-52}) (or
Eq.~(\ref{eq:kouchan-A2is1-53})) and
Eq.~(\ref{eq:Pb1-Pb2_in_ISTKH2011-0}) look similar, but their
identification is still not confirmed.


In this paper, we discuss the consistency between the ISTKH
experiment and results in
Ref.~\cite{K.Nakamura-A.Nishizawa-M.-K.Fujimoto-2012}.
The essential difference between the analysis in
Ref.~\cite{K.Nakamura-A.Nishizawa-M.-K.Fujimoto-2012} and the
ISTKH experiment is in the treatment of the entangled state.
In the AAV scenario, the weakly entangled state is created
through the weak von-Neumann interaction, while the ISTKH
experiment produces the maximally entangled state of the system
and the measuring device.
To overcome this difference in the re-interpretation of the ISTKH, 
we choose the state just after the entanglement creation 
as a pre-selected state.
This choice gives rise to the other difficulty that we cannot
distinguish the pre-selected state of only the system, because
the entangled state involves both initial states of the
measuring device and the system.
To treat this state as the pre-selected state, we introduce the
notion of {\it extended weak values}
\begin{eqnarray}
  \label{eq:extended-weak-values-intro}
  \langle\hat{O}\otimes\hat{P}\rangle_{w(i)}
  :=
  \frac{
    \langle\psi_{f(i)}|\hat{O}\otimes\hat{P}|\psi_{comp}\rangle
  }{
    \langle\psi_{f(i)}|\psi_{comp}\rangle
  }
\end{eqnarray}
instead of the original definition of the weak value 
(\ref{eq:weak-value-def}), where $|\psi_{comp}\rangle$ is the 
pre-selected state of a composite system of the system and the measuring
device, $\hat{P}$ is an operator for the measuring device, and
$|\psi_{f(i)}\rangle$ is a product state of the post-selected state of the system 
and the final state of the measuring device. 
[See Eqs.~(\ref{eq:pre-selected-state}),
(\ref{eq:post-selected-states})--(\ref{eq:psi_f1_f2_psi_i_products}),
(\ref{eq:Nishizawa-weak-values-1}), and
(\ref{eq:Nishizawa-weak-values-2}).]
This extended definition of weak values is a natural extension
of the original definition (\ref{eq:weak-value-def}) and very useful 
when we equivalently treat the system and the measurement device.
In our case, $|\psi_{comp}\rangle$ is chosen as a maximally
entangled state.
Through these newly defined weak values, it becomes possible to 
re-interpret the experiments in ISTKH along the original AAV
scenario and to show that the results
Eq.~(\ref{eq:kouchan-A2is1-52}) (or
Eq.~(\ref{eq:kouchan-A2is1-53})) and
Eq.~(\ref{eq:Pb1-Pb2_in_ISTKH2011-0}) are consistent.
In the same manner, we also evaluated the fluctuations of the
results in ISTKH and consequently found that an optimal
measurement strength exists for a fixed weak value, on which the
fluctuations in the final measurement vanish.
We also discuss the physical meaning of this optimal condition 
and made a confirmation of the consistency with the experiment.


Organization of this paper is as follows:
In section~\ref{sec:Essence_of_the_experimental_setup}, we
briefly review the experimental setup in ISTKH.
In section~\ref{sec:Interpretations_of_the_experiment}, the ISTKH experiment 
is re-examined along the original scenario of
weak measurements proposed by AAV through the introduction of an
extended definition of weak values.
We show two different interpretations of this experiment: the
first one is a weak measurement with a real weak value; the
second one is a weak measurement with an imaginary weak value.
These two interpretations are essentially equivalent to each
other.
In section~\ref{sec:Fluctuations}, we discuss the behavior of
the fluctuations in the final measurement in ISTKH.
Final section (section~\ref{sec:Summary_and_Discussion}) is
devoted to summary and discussions which include discussions on
the consistency with experimental results.


Throughout this paper, we use the natural unit $\hbar=1$.


\section{Experimental setup in ISTKH}
\label{sec:Essence_of_the_experimental_setup}


\begin{figure}[ht]
  \centering
  \begin{center}
    \includegraphics[width=0.5\textwidth]{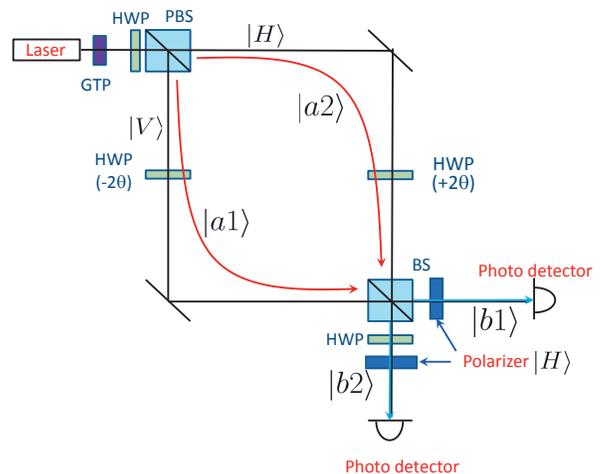}
  \end{center}
  \caption{
    Essential part in the experimental setup in the ISTKH
    experiment~\cite{M.Iinuma-Y.Suzuki-G.Taguchi-Y.Kadoya-H.F.Hofmann-2011}.
  }
  \label{fig:essence_of_ISTKH2011_setup}
\end{figure}


Here, we briefly summarize the experimental setup in ISTKH.
The details of the setup is given in ISTKH, but we restrict the
explanation to only significant points in the experiment.
The essential part of this experiment setup is depicted in
Figure~\ref{fig:essence_of_ISTKH2011_setup}.


In ISTKH, the observable in the system is the polarization of
photons and the pointer in the measuring device is the
which-path information.
As shown in Figure~\ref{fig:essence_of_ISTKH2011_setup}, the
incident photons from the laser go through the Glan-Thompson
prism (GTP) and the half-wave plate (HWP) before the polarized
beam splitter (PBS).
The GTP and HWP can transmit the photon with the linear
polarization and control the polarization angle of the incident
photons.
Therefore, these optical components are used for preparation of
the initial pure state $|\psi_{i}\rangle $ given by
\begin{equation}
  \label{eq:init-state}
  |\psi_{i}\rangle = C_{V}|V\rangle+C_{H}|H\rangle,
\end{equation}
where $\left|C_{V}\right|^{2}+\left|C_{H}\right|^{2}=1$ and
$C_{V}$ and $C_{H}$ are restricted to real numbers in this paper.
Each polarization state $|H\rangle$ and $|V\rangle$ 
corresponds to the horizontal and the vertical polarization, 
respectively~\cite{notation-comments}.
The PBS separates the initial beam into two paths
$\{|a1\rangle,|a2\rangle\}$ and simultaneously also the photon
polarization into the horizontal state $|H\rangle$ and the
vertical state $|V\rangle$.
Thus, the PBS produces the entangled state of the photon polarization 
and the path as
\begin{eqnarray}
  \label{eq:pre-selected-state}
  |\psi_{ent}\rangle
  =
  C_{V}|V\rangle\otimes|a1\rangle
  +
  C_{H}|H\rangle\otimes|a2\rangle.
\end{eqnarray}


After the PBS, the photons go through the other HWP on each
path a1 and a2 inside of the interferometer.
In the path a1, where the polarization state just before the HWP
is $|V\rangle$, the polarization is rotated by $-2\theta$ to the
clockwise direction by the HWP.
On the other hand, in the path a2, where the polarization state
just before the HWP is $|H\rangle$, the polarization is rotated
by $2\theta$ to the counter-clockwise direction by the HWP.
These rotations of the polarizations are regarded as the weak
interaction in the weak measurement.


Each path a1 and a2 is overlapped at the beam
splitter (BS) with the 50:50 ratio of the transmittance to the
reflectance and the photon's path-state is transformed into the
states $|b1\rangle$ and $|b2\rangle$ as
\begin{eqnarray}
  \label{eq:b1b2-a1a2-relation}
  |b1\rangle
  =
  \frac{1}{\sqrt{2}} \left(
    |a1\rangle + |a2\rangle
  \right)
  , 
  |b2\rangle
  =
  \frac{1}{\sqrt{2}} \left(
    - |a1\rangle + |a2\rangle
  \right)
  .
\end{eqnarray}
The transformation (\ref{eq:b1b2-a1a2-relation}) of the Hilbert
space $\{|a1\rangle,|a2\rangle\}$ to the Hilbert space
$\{|b1\rangle,|b2\rangle\}$ is an unitary transformation which
satisfies
\begin{eqnarray}
  \label{eq:b1b2-a1a2-completeness}
  |a1\rangle\langle a1|
  +
  |a2\rangle\langle a2|
  =
  |b1\rangle\langle b1|
  +
  |b2\rangle\langle b2|
  =
  \hat{1}
  .
\end{eqnarray}


On the path b2 from the output port of BS, a HWP is installed to
disentangle the system and the measuring device.
After the BS, the photons go through a polarizer on each path b1
and b2 for the post-selection of the weak measurement.
In the ISTKH experiment, the final polarization state is
selected to the horizontal state $|H\rangle$ by these
polarizers.


The photons in each path b1 and b2 are detected by the single
photon detector on each path.
The number of photons 
are counted and
their results give the conditional probabilities associated with
the which-path information $|b1\rangle$ and $|b2\rangle$.
Here, we denote these probability distributions $P(b1)$ and
$P(b2)$, respectively.


Finally, the difference $P(b1)-P(b2)$ is calculated, because the
weak value is obtained by dividing an average of the pointer
variable with the obtained conditional probabilities by the
measurement strength.
Therefore, the difference of the conditional probabilities
$P(b1)-P(b2)$ can be considered as the final result in this
experiment.


According to ISTKH, the expression of $P(b1)-P(b2)$ is given by
\begin{eqnarray}
  P(b1) - P(b2)
  &=&
  \frac{
    \sin(4\theta) \mbox{Re}\langle\hat{\sigma}_{x}\rangle_{w}
  }{
    1
    +
    \frac{1}{2} \left(
      1 - \left|\langle\hat{\sigma}_{x}\rangle_{w}\right|^{2}
    \right)
    \left(
      \cos(4\theta) - 1
    \right)
  }
  ,
  \nonumber\\
  &&
  \label{eq:Pb1-Pb2_in_ISTKH2011-2}
\end{eqnarray}
where $\langle\hat{\sigma}_{x}\rangle_{w}=C_{V}/C_{H}$.
Eq.~(\ref{eq:Pb1-Pb2_in_ISTKH2011-2}) is the precise expression
of Eq.~(\ref{eq:Pb1-Pb2_in_ISTKH2011-0}), which was derived from
the POVM formalism and the combination of unitary
transformations.
One of the purposes of this paper is re-derivation of
Eq.~(\ref{eq:Pb1-Pb2_in_ISTKH2011-2}) along the original
scenario by AAV~\cite{Y.Aharonov-D.Z.Albert-L.Vaidman-1988}.


\section{Interpretations of the experiment in ISTKH through AAV scenario}
\label{sec:Interpretations_of_the_experiment}


In the original AAV
argument~\cite{Y.Aharonov-D.Z.Albert-L.Vaidman-1988}, the weak
measurement consists of four processes, pre-selection, weak
interaction, post-selection, and the final measurement of the
measuring device.
However, these serial processes do not explicitly include the
entanglement creation.
Therefore, the correspondence between a sequence of the above
processes and an alternative sequence in the ISTKH experiment
remains unclear.


One of ways for excluding the entanglement creation from the whole
process of the weak measurement is to use the state
Eq.~(\ref{eq:pre-selected-state}) as the pre-selected state
instead of Eq.~(\ref{eq:init-state}).
This requires a mathematically equivalent treatment for the
device's state as the system's one.
In this paper, we resolved this problem by the introduction of
two {\it extended weak values} associated also with the final
state of the measuring device, which are defined in later
(Eqs.~(\ref{eq:Iinuma-weak-values-1}) and
(\ref{eq:Iinuma-weak-values-2}) in section
\ref{sec:Interpretations_of_the_experiment_real_sigmax} or
Eqs.~(\ref{eq:Nishizawa-weak-values-1}) and
(\ref{eq:Nishizawa-weak-values-2}) in section
\ref{sec:Interpretations_of_the_experiment_imaginary_sigmay}).


On the process of the post-selection, the final state
$|H\rangle$ is selected by the polarizers just before the photon
detection in the ISTKH experiment.
It is therefore straightforward to treat the photon polarization
with the basis of $\{|H\rangle,|V\rangle\}$ as the ``system''.
For the selection of $|H\rangle$ as the final state, the effect
of the HWP installed just after the BS does not consequently
affect the final result $P(b1)-P(b2)$, because this HWP just
gives the $\pi$ phase difference between the state $|H\rangle$
and the state $|V\rangle$.
This final result can be obtained by the final measurement of
the ``measuring device'', that is the projection measurement to
the basis $\{|b1\rangle,|b2\rangle\}$ representing which-path
information.
This is easily understood by rewriting the probability
distribution $P(b1)-P(b2)$ in the form 
\begin{eqnarray}
  P(b1)-P(b2)
  &:=&
  \langle b1|\rho_{f}|b1\rangle
  -
  \langle b2|\rho_{f}|b2\rangle
  \nonumber\\
  &=&
  \mbox{Tr}\left[
    \left(
      |b1\rangle\langle b1|
      -
      |b2\rangle\langle b2|
    \right)\rho_{f}
  \right]
  ,
\end{eqnarray}
where $\rho_{f}$ is the conditional density matrix after the
post-selection.
The above equation shows that $P(b1)-P(b2)$ is the expectation
value of the which-path operator 
$|b1\rangle\langle b1|-|b2\rangle\langle b2|$ and the final
measurement corresponds to the projection measurement of 
the Hilbert space $\{|b1\rangle,|b2\rangle\}$.


The remaining correspondence is on the process of the weak
interaction between the system and the measuring device. 
The identification of the interaction Hamiltonian in the ISTKH
experiment is necessary to understand the full process along the
AAV's scenario.
We found that two effective interaction Hamiltonians, which
provide the explanation of the ISTKH experiment in a different
but essentially equivalent way.
One is the interpretation by a real extended weak value for the
operator $\hat{\sigma}_{x}$ and another is by an imaginary
extended weak value for $\hat{\sigma}_{y}$.
These two interpretations give the essentially same results.
Thus, both of these interpretations reproduce the result of the
ISTKH experiment along the original proposal of
AAV~\cite{Y.Aharonov-D.Z.Albert-L.Vaidman-1988}.


\subsection{Interpretation with a real extended weak value for $\hat{\sigma}_{x}$}
\label{sec:Interpretations_of_the_experiment_real_sigmax}


The polarization in each path a1 and a2 is rotated in opposite
direction by the HWPs inside the interferometer. 
As the effective Hamiltonian describing the interaction between
the polarization and the which-path operator, we consider the
following form:
\begin{eqnarray}
  \label{eq:Iinuma-HWP-interaction}
  \hat{H}_{HWP}
  =
  \Omega \left(
    \sin(2\theta)\frac{\hat{\sigma}_{x}}{2}\otimes\hat{A}
    +
    \cos(2\theta)\frac{\hat{\sigma}_{z}}{2}\otimes\hat{1}
  \right),
\end{eqnarray}
where $\theta$ is an angle of the fast axis of HWP from the
horizontal axis and $\Omega\Delta t$ represents a phase
difference between the fast component and the slow component 
after passing through the HWP.
In the case of HWP, the phase difference is $\Omega\Delta t=\pi$
with $\Delta t=l/c$, where $l$ is the length of the HWP.
The operator $\hat{A}$ is defined by
\begin{eqnarray}
  \label{eq:a1-a2-which-path-operator}
  \hat{A} := - |a1\rangle\langle a1| + |a2\rangle\langle a2|,
\end{eqnarray}
which has an eigenvalue of $\pm 1$ for each path eigenstate of
$|a1\rangle$ and $|a2\rangle$.
Therefore, the which-path operator determines a sign of the
first term including $\hat{\sigma}_{x}$ in
Eq.~(\ref{eq:Iinuma-HWP-interaction}).


From the interaction Hamiltonian
Eq.~(\ref{eq:Iinuma-HWP-interaction}), the evolution operator
$\hat{U}_{1}$ through this interaction is given by 
\begin{eqnarray}
  \hat{U}_{1}
  &=&
  \exp\left[
    - i \hat{H}_{HWP} \Delta t
  \right]
  \nonumber\\
  &=&
  \exp\left[
    -i\frac{\pi}{2} \left(
      \sin(2\theta)\hat{\sigma}_{x}\otimes\hat{A}
      +
      \cos(2\theta)\hat{\sigma}_{z}\otimes\hat{1}
    \right)
  \right]
  .
  \nonumber\\
  &&
\end{eqnarray}
Here, we note that 
\begin{eqnarray}
  \left(
    \sin(2\theta)\hat{\sigma}_{x}\otimes\hat{A}
    +
    \cos(2\theta)\hat{\sigma}_{z}\otimes\hat{1}
  \right)^{2}
  =
  \hat{1}\otimes\hat{1}
  .
\end{eqnarray}
From this property, the unitary operator $\hat{U}_{1}$ can be 
written as 
\begin{eqnarray}
  \label{eq:Iinuma-HWP-evolution-operator}
  \hat{U}_{1}
  &=&
  - i \left(
    \sin(2\theta)\hat{\sigma}_{x}\otimes\hat{A}
    +
    \cos(2\theta)\hat{\sigma}_{z}\otimes\hat{1}
  \right)
  . 
\end{eqnarray}


This unitary operator transforms the polarization vector into in
the line-symmetric position with respect to the fast axis of the
HWP, which angle is  $-\theta$ from the vertical axis in path a1
or $+\theta$ from the horizontal axis in path a2.
As the result, these transformations correspond to the effective
rotation of the polarization in a1 with the angle of $-2\theta$ 
from the vertical axis and of the polarization in a2 with the
angle of $+2\theta$ from the horizontal axis.
If the which-path operator $\hat{A}$ is absent, the unitary operator
(\ref{eq:Iinuma-HWP-evolution-operator}) becomes a well-known form 
for HWP~\cite{Yariv97}.
By the unitary transformation
(\ref{eq:Iinuma-HWP-evolution-operator}), the total density
matrix is evolved from the initial density matrix
$\rho_{init}=|\psi_{ent}\rangle\langle\psi_{ent}|$ to
$\hat{U}_{1}\rho_{in}\hat{U}^{\dagger}_{1}$, where the initial
state $|\psi_{ent}\rangle$ is given by
Eq.~(\ref{eq:pre-selected-state}).


Following this unitary evolution, the 50:50 BS transforms the basis
of which-path information $\{|a1\rangle,|a2\rangle\}$ into the
other basis $\{|b1\rangle,|b2\rangle\}$. 
Since this transformation can be written as the multiplication
of the identity operator (\ref{eq:b1b2-a1a2-completeness}), we
can consider this BS transformation after the post-selection of
the system.


After selecting to the final state $|H\rangle$ by the polarizers 
downstream of the BS, the conditional density matrix can be expressed as  
\begin{eqnarray}
  \label{eq:Iinuma-final-density-matrix}
  \rho_{f}
  &=&
  \frac{|\psi_{f}\rangle\langle\psi_{f}|}{\langle\psi_{f}|\psi_{f}\rangle}
  , \\
  |\psi_{f}\rangle
  &=&
  - i \langle H|\otimes\hat{1}\left[
    \sin(2\theta)\hat{\sigma}_{x}\otimes\hat{A}
  \right.
  \nonumber\\
  && \quad\quad\quad\quad\quad
  \left.
    +
    \cos(2\theta)\hat{\sigma}_{z}\otimes\hat{1}
  \right]|\psi_{ent}\rangle,
  \label{eq:Iinuma-final-state}
\end{eqnarray}
where the identity $\hat{1}$ in
Eq.~(\ref{eq:Iinuma-final-state}) is given by
(\ref{eq:b1b2-a1a2-completeness}).


To treat the post-selected state of the system and the projected
states of the measuring device equivalently, we introduce
following two final states 
\begin{eqnarray}
  \label{eq:post-selected-states}
  |\psi_{f(1)}\rangle := |H\rangle\otimes|b1\rangle, \quad
  |\psi_{f(2)}\rangle := |H\rangle\otimes|b2\rangle.
\end{eqnarray}
Further, we also introduce two {\it extended weak values} which
are defined by
\begin{eqnarray}
  \label{eq:Iinuma-weak-values-1}
  \left\langle\hat{\sigma}_{x}\otimes\hat{A}\right\rangle_{w(1)}
  &:=&
  \frac{
    \langle\psi_{f(1)}|\hat{\sigma}_{x}\otimes\hat{A}|\psi_{ent}\rangle
  }{
    \langle\psi_{f(1)}|\psi_{ent}\rangle
  }
  =
  - \frac{C_{V}}{C_{H}}
  , \\
  \label{eq:Iinuma-weak-values-2}
  \left\langle\hat{\sigma}_{x}\otimes\hat{A}\right\rangle_{w(2)}
  &:=&
  \frac{
    \langle\psi_{f(2)}|\hat{\sigma}_{x}\otimes\hat{A}|\psi_{ent}\rangle
  }{
    \langle\psi_{f(2)}|\psi_{ent}\rangle
  }
  =
  \frac{C_{V}}{C_{H}}
  .
\end{eqnarray}
The above weak values have real values.
In addition, the following relations are satisfied,
\begin{eqnarray}
  \label{eq:Iinuma-weak-values-sigmaz}
  \frac{
    \langle\psi_{f(1)}|\hat{\sigma}_{z}\otimes\hat{1}|\psi_{ent}\rangle
  }{
    \langle\psi_{f(1)}|\psi_{ent}\rangle
  }
  =
  \frac{
    \langle\psi_{f(2)}|\hat{\sigma}_{z}\otimes\hat{1}|\psi_{ent}\rangle
  }{
    \langle\psi_{f(2)}|\psi_{ent}\rangle
  }
  =
  1,
  \\
  \label{eq:psi_f1_f2_psi_i_products}
  \langle\psi_{f(1)}|\psi_{ent}\rangle
  =
  \langle\psi_{f(2)}|\psi_{ent}\rangle
  =
  \frac{1}{\sqrt{2}} C_{H}
  .
\end{eqnarray}
By using 
Eq.~(\ref{eq:Iinuma-weak-values-1})--(\ref{eq:psi_f1_f2_psi_i_products}),  
the normalization factor $\langle\psi_{f}|\psi_{f}\rangle$ of
the final density matrix
Eq.~(\ref{eq:Iinuma-final-density-matrix}) is expressed
as~\cite{signature-comments} 
\begin{eqnarray}
  \langle\psi_{f}|\psi_{f}\rangle
  &=&
  2 \left|\langle\psi_{ent}|\psi_{f(1)}\rangle\right|^{2}
  \left[
    \cos^{2}(2\theta)
  \right.
  \nonumber\\
  && \quad\quad\quad\quad\quad\quad\quad
  \left.
    +
    \sin^{2}(2\theta)
    \left|
      \left\langle\hat{\sigma}_{x}\otimes\hat{A}\right\rangle_{w(1)}
    \right|^{2}
  \right]
  .
  \nonumber\\
  &&
  \label{eq:Iinuma-psi_f-normalization}
\end{eqnarray}


The difference of probability distribution $P(b1)-P(b2)$
can be evaluated from
\begin{eqnarray}
  P(b1)-P(b2)
  &=&
  \langle b1|\rho_{f}|b1\rangle
  - 
  \langle b2|\rho_{f}|b2\rangle
  \nonumber\\
  &=&
  \frac{
    \left|\langle b1|\psi_{f}\rangle\right|^{2}
    -
    \left|\langle b2|\psi_{f}\rangle\right|^{2}
  }{
    \langle\psi_{f}|\psi_{f}\rangle
  }
  \label{eq:Pb1-Pb2-def}
\end{eqnarray}
\begin{widetext}
and
\begin{eqnarray}
  \label{eq:Iinuma-b1-psif-sigmax}
  \left|\langle b1|\psi_{f}\rangle\right|^{2}
  &=&
  \left|\langle\psi_{f(1)}|\psi_{ent}\rangle\right|^{2}
  \left\{
    \left|\left\langle\hat{\sigma}_{x}\otimes\hat{A}\right\rangle_{w(1)}\right|^{2}
    \sin^{2}(2\theta) 
    + \cos^{2}(2\theta)
    + \mbox{Re}\left\langle\hat{\sigma}_{x}\otimes\hat{A}\right\rangle_{w(1)}
    \sin(4\theta)
  \right\}
  , \\
  \label{eq:Iinuma-b2-psif-sigmax}
  \left|\langle b2|\psi_{f}\rangle\right|^{2}
  &=&
  \left|\langle\psi_{f(1)}|\psi_{ent}\rangle\right|^{2}
  \left\{
    \left|\left\langle\hat{\sigma}_{x}\otimes\hat{A}\right\rangle_{w(1)}\right|^{2}
    \sin^{2}(2\theta) 
    + \cos^{2}(2\theta)
    - \mbox{Re}\left\langle\hat{\sigma}_{x}\otimes\hat{A}\right\rangle_{w(1)}
    \sin(4\theta)
  \right\}
  .
\end{eqnarray}
Using
Eq.~(\ref{eq:Iinuma-final-density-matrix})--(\ref{eq:Iinuma-b2-psif-sigmax}),
we obtain the expression of $P(b1)-P(b2)$ as
\begin{eqnarray}
  \label{eq:Iinuma-Pb1-Pb2-results}
  P(b1) - P(b2)
  &=&
  \frac{
    \mbox{Re}\left\langle\hat{\sigma}_{x}\otimes\hat{A}\right\rangle_{w(1)}
    \sin(4\theta)
  }{
    1
    +
    \frac{1}{2} \left(
      1
      -
      \left|\left\langle\hat{\sigma}_{x}\otimes\hat{A}\right\rangle_{w(1)}\right|^{2}
    \right)
    \left(
      \cos(4\theta) - 1
    \right)
  }
  ,
\end{eqnarray}
\end{widetext}
which is essentially equivalent to
Eq.~(\ref{eq:Pb1-Pb2_in_ISTKH2011-2}) replacing the extended weak values
with the conventional weak values.
The appearance of the which-path operator $\hat{A}$ in the
definition of the extended weak value is a natural consequence
of our extension of their definitions 
(\ref{eq:Iinuma-weak-values-1}) and
(\ref{eq:Iinuma-weak-values-2}).
The result of Eq.~(\ref{eq:Iinuma-Pb1-Pb2-results}) shows that
the ISTKH experiment can be regard as the measurement of the
real extended weak value for the operator $\hat{\sigma}_{x}$ as
commented in
Ref.~\cite{K.Nakamura-A.Nishizawa-M.-K.Fujimoto-2012}.


\subsection{Interpretation with an imaginary weak value for $\hat{\sigma}_{y}$}
\label{sec:Interpretations_of_the_experiment_imaginary_sigmay}


The effective rotation of the photon polarization by the HWP
in the path a2 is also described by the active transformation of
$\{C_{H},C_{V}\}$ as
\begin{eqnarray}
  \left[
    \begin{array}{c}
      C_{H} \\
      C_{V}
    \end{array}
  \right]
  &\rightarrow&
  \left[
    \begin{array}{rr}
      \cos(2\theta) & - \sin(2\theta) \\
      \sin(2\theta) & \cos(2\theta)
    \end{array}
  \right]
  \left[
    \begin{array}{c}
      C_{H} \\
      C_{V}
    \end{array}
  \right]
  \nonumber\\
  &=&
  \left\{
    \left[
      \begin{array}{rr}
        1 & 0 \\
        0 & 1
      \end{array}
    \right]
    -
    \left[
      \begin{array}{cc}
        0 & 2\theta \\
        - 2\theta & 0
      \end{array}
    \right]
    + O(\theta^{2})
  \right\}
    \left[
    \begin{array}{c}
      C_{H} \\
      C_{V}
    \end{array}
  \right].
  \nonumber\\
  &&
  \label{eq:active_rotation_in_CH-CV_plane}
\end{eqnarray}
Since the unitary operator generated by an operator $\hat{G}$ 
is expressed as
$U(\theta)=e^{-i\alpha\hat{G}}=1-i\alpha\hat{G}+O(\alpha^{2})$,
the rotation expressed as Eq.~(\ref{eq:active_rotation_in_CH-CV_plane}) 
is generated by the operator 
$\hat{G}=\hat{\sigma}_{y}=-i|H\rangle\langle V|+i|V\rangle\langle H|$ 
with the parameter $\alpha=2\theta$.
In the path a1, the polarization can be also rotated as
Eq.~(\ref{eq:active_rotation_in_CH-CV_plane}) with the opposite
direction ($\theta\rightarrow-\theta$).
Using the which-path operator $\hat{A}$ defined by
Eq.~(\ref{eq:a1-a2-which-path-operator}), therefore,
the unitary evolution by the HWP on each path a1 and a2 is
written as
\begin{eqnarray}
  \label{eq:Nishizawa-unitary-operator-0}
  \hat{U}_{2}
  :=
  \exp\left[-i(2\theta)\hat{\sigma}_{y}\otimes\hat{A}\right]
  . 
\end{eqnarray}
The effective interaction Hamiltonian can be expressed as 
\begin{eqnarray}
  \label{eq:Nishizawa-HWP-interaction}
  \hat{H}_{2} = (2\theta)\hat{\sigma}_{y}\otimes\hat{A}
  \delta(t-t_{0}). 
\end{eqnarray}
Since the operator $\hat{\sigma}_{y}\otimes\hat{A}$
satisfies the property 
$\left(\hat{\sigma}_{y}\otimes\hat{A}\right)^{2}=\hat{1}\otimes\hat{1}$,
the unitary operator Eq.~(\ref{eq:Nishizawa-unitary-operator-0})
is also represented by
\begin{eqnarray}
  \label{eq:Nishizawa-unitary-operator-1}
  \hat{U}_{2}
  =
  \cos(2\theta) \hat{1}\otimes\hat{1}
  -
  i \sin(2\theta) \hat{\sigma}_{y}\otimes\hat{A}
  .
\end{eqnarray}
This unitary operator describes the rotation of photon
polarizations on each path a1 and a2.
Therefore, instead of the operator $\hat{U}_{1}$ given by
Eq.~(\ref{eq:Iinuma-HWP-evolution-operator}), we can use the
operator $\hat{U}_{2}$ given by
Eq.~(\ref{eq:Nishizawa-unitary-operator-1}).
This unitary evolution transforms the conditional density matrix
from the initial density matrix
$\rho_{init}=|\psi_{ent}\rangle\langle\psi_{ent}|$ to
$\hat{U}_{2}\rho_{in}\hat{U}^{\dagger}_{2}$,
where the pre-selected state $|\psi_{ent}\rangle$ is given by
Eq.~(\ref{eq:pre-selected-state}).


In the same way as the AAV scenario, the conditional
density matrix $\rho_{f}$ after the post-selection can be written as
Eq.~(\ref{eq:Iinuma-final-density-matrix}) with 
\begin{eqnarray}
  |\psi_{f}\rangle
  &=&
  \langle H|\otimes\hat{1} \left[
    \cos(2\theta)\hat{1}\otimes\hat{1}
    -
    i \sin(2\theta) \hat{\sigma}_{y}\otimes\hat{A} 
  \right]|\psi_{ent}\rangle
  \nonumber\\
  &&
  \label{eq:Nishizawa-final-state}
\end{eqnarray}
instead of Eq.~(\ref{eq:Iinuma-final-state}).


In the same manner in section
\ref{sec:Interpretations_of_the_experiment_real_sigmax}, 
we introduce two final states given by
Eq.~(\ref{eq:post-selected-states}) and two {\it extended weak
  values} which are defined by
\begin{eqnarray}
  \label{eq:Nishizawa-weak-values-1}
  \left\langle\hat{\sigma}_{y}\otimes\hat{A}\right\rangle_{w(1)}
  \!&:=&\!
  \frac{
    \langle\psi_{f(1)}|\hat{\sigma}_{y}\otimes\hat{A}|\psi_{ent}\rangle
  }{
    \langle\psi_{f(1)}|\psi_{ent}\rangle
  }
  =
  i \frac{C_{V}}{C_{H}}
  , \\
  \label{eq:Nishizawa-weak-values-2}
  \left\langle\hat{\sigma}_{y}\otimes\hat{A}\right\rangle_{w(2)}
  \!&:=&\!
  \frac{
    \langle\psi_{f(2)}|\hat{\sigma}_{y}\otimes\hat{A}|\psi_{ent}\rangle
  }{
    \langle\psi_{f(2)}|\psi_{ent}\rangle
  }
  =
  - i \frac{C_{V}}{C_{H}}
  .
\end{eqnarray}
In contrast to the argument in section
\ref{sec:Interpretations_of_the_experiment_real_sigmax}, these
weak values are imaginary.
By using Eq.~(\ref{eq:psi_f1_f2_psi_i_products}) and
Eq.~(\ref{eq:Nishizawa-final-state})--(\ref{eq:Nishizawa-weak-values-2}),
the normalization factor $\langle\psi_{f}|\psi_{f}\rangle$ of
the final density matrix can be written as 
\begin{eqnarray}
  \langle\psi_{f}|\psi_{f}\rangle
  &=&
  2 \left|\langle\psi_{ent}|\psi_{f(1)}\rangle\right|^{2}
  \left[
    \cos^{2}(2\theta)
  \right.
  \nonumber\\
  && \quad\quad\quad\quad\quad\quad\quad
  \left.
    +
    \sin^{2}(2\theta)
    \left|
      \left\langle\hat{\sigma}_{y}\otimes\hat{A}\right\rangle_{w(1)}
    \right|^{2}
  \right]
  .
  \nonumber\\
  &&
  \label{eq:Nishizawa-psi_f-normalization}
\end{eqnarray}


\begin{widetext}
The probability difference $P(b1)-P(b2)$ can be evaluated with
Eq.~(\ref{eq:Pb1-Pb2-def}).
From Eq.~(\ref{eq:psi_f1_f2_psi_i_products}),
and
Eq.~(\ref{eq:Nishizawa-final-state})--(\ref{eq:Nishizawa-weak-values-2}),
we get 
\begin{eqnarray}
  \label{eq:Iinuma-b1-psif-sigmay}
  \left|\langle b1|\psi_{f}\rangle\right|^{2}
  &=&
  \left|\langle\psi_{f(1)}|\psi_{ent}\rangle\right|^{2}
  \left\{
    \left|\left\langle\hat{\sigma}_{y}\otimes\hat{A}\right\rangle_{w(1)}\right|^{2}
    \sin^{2}(2\theta) 
    + \cos^{2}(2\theta)
    - \mbox{Im}\left\langle\hat{\sigma}_{y}\otimes\hat{A}\right\rangle_{w(1)}
    \sin(4\theta)
  \right\}
  , \\
  \label{eq:Iinuma-b2-psif-sigmay}
  \left|\langle b2|\psi_{f}\rangle\right|^{2}
  &=&
  \left|\langle\psi_{f(1)}|\psi_{ent}\rangle\right|^{2}
  \left\{
    \left|\left\langle\hat{\sigma}_{y}\otimes\hat{A}\right\rangle_{w(1)}\right|^{2}
    \sin^{2}(2\theta) 
    + \cos^{2}(2\theta)
    + \mbox{Im}\left\langle\hat{\sigma}_{y}\otimes\hat{A}\right\rangle_{w(1)}
    \sin(4\theta)
  \right\}
  .
\end{eqnarray}
Thus, the expression of $P(b1)-P(b2)$ can be obtained
as~\cite{signature-comments}  
\begin{eqnarray}
  \label{eq:Nishizawa-Pb1-Pb2-results}
  P(b1) - P(b2)
  &=&
  \frac{
    -
    \mbox{Im}\left\langle\hat{\sigma}_{y}\otimes\hat{A}\right\rangle_{w(1)}
    \sin(4\theta)
  }{
    1
    +
    \frac{1}{2} \left(
      1
      -
      \left|\left\langle\hat{\sigma}_{y}\otimes\hat{A}\right\rangle_{w(1)}\right|^{2}
    \right)
    \left(
      \cos(4\theta) - 1
    \right)
  }
  ,
\end{eqnarray}
\end{widetext}
which coincides with Eq.~(\ref{eq:Pb1-Pb2_in_ISTKH2011-2}) by
the replacement of $\mbox{Re}\langle\hat{\sigma}_{x}\rangle_{w}$ 
$\rightarrow$
$\mbox{Im}\langle\hat{\sigma}_{y}\otimes\hat{A}\rangle_{w(1)}$.
The appearance of the which-path operator $\hat{A}$ 
is a natural consequence of our extension of weak values.
It is obvious that the result
Eq.~(\ref{eq:Nishizawa-Pb1-Pb2-results}) corresponds to
Eq.~(\ref{eq:kouchan-A2is1-53}) which was derived
in~\cite{K.Nakamura-A.Nishizawa-M.-K.Fujimoto-2012}.
Eq.~(\ref{eq:Iinuma-Pb1-Pb2-results}) and
Eq.~(\ref{eq:Nishizawa-Pb1-Pb2-results}) show that the ISTKH
experiment can be interpreted not only as a weak measurement
of a real weak value but also that of an imaginary weak value.


\section{Fluctuations}
\label{sec:Fluctuations}


In the previous section, we showed that the two expectation  
values $P(b1)-P(b2)$ derived from the two different
interpretations are the essentially same.
In this section, we discuss the fluctuations in $P(b1)-P(b2)$.
Since $P(b1)-P(b2)$ is the expectation value of the which-path
operator $|b1\rangle\langle b1|-|b2\rangle\langle b2|$, 
the corresponding variance is given by
\begin{eqnarray}
  &&
  \mbox{Var}\left(
    |b1\rangle\langle b1|
    -
    |b2\rangle\langle b2|
  \right)
  \nonumber\\
  &:=&
  \mbox{Tr}\left[
    \left(
      |b1\rangle\langle b1|
      -
      |b2\rangle\langle b2|
    \right)^{2}
    \rho_{f}
  \right]
  \nonumber\\
  &&
  -
  \left(
    \mbox{Tr}\left[
      \left(
        |b1\rangle\langle b1|
        -
        |b2\rangle\langle b2|
      \right)
      \rho_{f}
    \right]
  \right)^{2}
  .
  \label{eq:variance-def}
\end{eqnarray}
Using the property 
$\left(|b1\rangle\langle b1|-|b2\rangle\langle b2|\right)^{2}=\hat{1}$,
Eq. (\ref{eq:variance-def}) provides the same variance 
for the same value $P(b1)-P(b2)$.
Therefore, we show the fluctuations in $P(b1)-P(b2)$ obtained
only from the interpretation in section
\ref{sec:Interpretations_of_the_experiment_real_sigmax}.


Using Eq. (\ref{eq:Iinuma-Pb1-Pb2-results}), we can easily obtain 
\begin{eqnarray}
  \mbox{Var}\left(
    |b1\rangle\langle b1|
    -
    |b2\rangle\langle b2|
  \right)
  &=&
  \left(
    \frac{
      1
      -
      \zeta^{2}
    }{
      1
      +
      \zeta^{2}
    }
  \right)^{2}
  ,
\end{eqnarray}
where we used
$\mbox{Im}\left\langle\hat{\sigma}_{x}\otimes\hat{A}\right\rangle_{w(1)}=0$
and defined $\zeta$ by
\begin{eqnarray}
  \label{eq:zeta-def}
  \zeta
  := 
  \tan(2\theta)
  \mbox{Re}\left\langle\hat{\sigma}_{x}\otimes\hat{A}\right\rangle_{w(1)}. 
\end{eqnarray}
Thus, the fluctuation
$\Delta(P(b1)-P(b2)):=\sqrt{\mbox{Var}(|b1\rangle\langle
  b1|-|b2\rangle\langle b2|)}$ is given by 
\begin{eqnarray}
  \label{eq:Nakamura-fluctuation-results}
  \Delta(P(b1)-P(b2))
  &=&
  \left|
    \frac{
      1
      -
      \zeta^{2}
    }{
      1
      +
      \zeta^{2}
    }
  \right|
  .
\end{eqnarray}
It is obvious that the fluctuations expressed by
Eq.~(\ref{eq:Nakamura-fluctuation-results}) vanishes when the
condition 
\begin{eqnarray}
  \label{eq:Nakamura-optimal-point}
  \zeta
  = 
  \pm 1
\end{eqnarray}
is satisfies.
We will discuss the physical meaning of these optimal conditions
later.


In the following, we discuss the behaviors of
$P(b1)-P(b2)$ and its fluctuation $\Delta(P(b1)-P(b2))$.
The difference $P(b1)-P(b2)$ given by
Eq.~(\ref{eq:Iinuma-Pb1-Pb2-results}) can be rewritten as the
form 
\begin{eqnarray}
  \label{eq:Nakamura-Pb1-Pb2-expression}
  P(b1) - P(b2)
  &=&
  \frac{
    2 \zeta
  }{
    1
    +
    \zeta^{2}
  }
  ,
\end{eqnarray}
where we used
$\mbox{Im}\left\langle\hat{\sigma}_{x}\otimes\hat{A}\right\rangle_{w(1)}=0$. 
When $|\zeta|>>1$, $P(b1)-P(b2)\sim 2/\zeta$ and 
$\Delta(P(b1)-P(b2))\sim 1$ are given from
Eq.~(\ref{eq:Nakamura-fluctuation-results}) and 
Eq.~(\ref{eq:Nakamura-Pb1-Pb2-expression}). 
Therefore, in the regime $|\zeta|>>1$ the fluctuation
dominates and clearly makes it difficult to measure the signal 
$P(b1)-P(b2)$.
On the other hands, at $\zeta=0$ and $|\zeta|=\infty$,
$P(b1)-P(b2)=0$ is satisfied.
This leads to the presence of extremal points in $P(b1)-P(b2)$.
Differentiating $P(b1)-P(b2)$ with respect to $\zeta$, we can
easily confirm that these extremal points coincide with the
optimal conditions Eq.~(\ref{eq:Nakamura-optimal-point}), i.e.,
$P(b1)-P(b2)=1$ at $\zeta=1$ and $P(b1)-P(b2)=-1$ at $\zeta=-1$.


Since $0\leq P(b1),P(b2)\leq1$ is satisfied, each relation of
$P(b1)-P(b2)=1$ and $P(b1)-P(b2)=-1$ results in
$(P(b1),P(b2))=(1,0)$ and $(P(b1),P(b2))=(0,1)$, respectively.
This indicates that the state
$|\psi_{f}\rangle/\sqrt{\langle\psi_{f}|\psi_{f}\rangle}$ of the
total system just after the post-selection should be an
eigenstate of the which-path operator
$|b1\rangle\langle b1|-|b2\rangle\langle b2|$
in the optimal condition Eq.~(\ref{eq:Nakamura-optimal-point}).
We can easily see it by rewriting the state $|\psi_{f}\rangle$
as follows:
\begin{eqnarray}
  |\psi_{f}\rangle
  &=&
  - i\cos(2\theta)\langle\psi_{f(1)}|\psi_{ent}\rangle
  \left\{
    \left(1+\zeta\right)|b1\rangle
  \right.
  \nonumber\\
  && \quad\quad\quad\quad\quad\quad\quad\quad\quad\quad
  \left.
    +
    \left(1-\zeta\right)|b2\rangle
  \right\}
  .
  \label{eq:Nakamura-psif-evaluation}
\end{eqnarray}
This expression clearly shows $\rho_{f}=|b1\rangle\langle b1|$
at $\zeta=1$ and $\rho_{f}=|b2\rangle\langle b2|$ at $\zeta=-1$,
respectively.
Thus, the measurement of the which-path observable
in the optimal conditions Eq.~(\ref{eq:Nakamura-optimal-point})
results in the measurement of the eigenstates of the operator
$|b1\rangle\langle b1|-|b2\rangle\langle b2|$ and the
fluctuations in $P(b1)-P(b2)$ vanish.


\section{Summary and Discussions}
\label{sec:Summary_and_Discussion}


In summary, we showed re-interpretations of the ISTKH
experiment on the back-action in a weak measurement through 
the AAV original scenario which was discussed in 
Ref.~\cite{K.Nakamura-A.Nishizawa-M.-K.Fujimoto-2012}.
Although the ISTKH experiment is different from the AAV argument
due to the explicit entanglement creation between the system and
the measuring device, the introduction of the extended weak
value makes us understand the ISTKH experiment along the
original scenario proposed by AAV, i.e., pre-selection, weak
interaction, post-selection, and the final measurement of the
measuring device.
The extended weak value is generally applicable to other experiments
requiring the mathematically equivalent treatment for the device
and the system.
As the result, we achieved two re-interpretations to the ISTKH
experiment, which correspond to two types of weak measurements:
one is for the real extended weak value and another is for the
imaginary extended weak value, but these re-interpretations give
the essentially same results.
These show that these two interpretations are essentially 
equivalent and the ISTKH experiment is consistent with the
original AAV argument and the results in
Ref.~\cite{K.Nakamura-A.Nishizawa-M.-K.Fujimoto-2012}.


The equivalence of two re-interpretation is very interesting,
because the usual argument on weak measurement~\cite{Jozsa07}
shows that the real and imaginary parts of weak value emerges as
the shift in the average of the pointer variable and its
conjugate momentum, respectively.
Unlike this argument, the ISTKH experiment is the first example
of the experiment which can be regarded as not only the
measurement of a real weak value, but also that of an imaginary
weak values at the same time.
This interesting feature is obviously provided by the properties
of Pauli operators.
Therefore, the similar weak measurements of qubit systems have a
possibility of including the same feature.
However, many experiments of qubit systems do not need a
knowledge of the explicit form of the effective Hamiltonian in
the aspect of practical implementations, so that there have been
no such arguments yet.


In addition, we also showed that the fluctuations in the
expectation value for both interpretations result in the
completely same formula.
We found the optimal condition in the measurement strength
$\theta$, where the fluctuations vanish for a fixed weak value.
This vanishing fluctuations do not certainly provide the
infinity of the signal to noise ratio in the optimal condition
because of the presence of other noise sources  (for example,
shot noise, imperfection in the setup, and so on). 
However, this feature of the fluctuation in the optimal condition is 
potentially useful for the improvement of the signal to noise ratio. 
In the reference~\cite{A.Nishizawa-K.Nakamura-M.-K.Fujimoto-2012},
they discussed the precision of a phase measurement by the weak
measurement in the shot-noise limited interferometer and found
that the photo-detector dominantly produces the photon shot-noise 
which determines the signal to noise ratio.
In the ISTKH experiment, the state just after the post-selection 
may be identified to the eigenstates of the which-path operator 
in the final measurement, so that the fluctuation becomes very small.
This is a new theoretical result on fluctuations and predicts 
that there are in principle rooms of the improvement of the fluctuations 
in the ISTKH experiments.
The considaration in
Ref.~\cite{A.Nishizawa-K.Nakamura-M.-K.Fujimoto-2012} and in this
paper implies that the final measurement of the detector in weak
measurements is crucial when we want to improve the
signal-to-noise.


The behavior of these fluctuations is consistent with the
current experiment~\cite{Y.Suzuki-M.Iinuma-H.F.Hofmann-2012}.
In the similar condition to Ref.~\cite{Y.Suzuki-M.Iinuma-H.F.Hofmann-2012}, 
the above calculation gives the optimal condition of
$\theta=11.25^{\circ}$ for the vanishing fluctuation.
From Ref.~\cite{Y.Suzuki-M.Iinuma-H.F.Hofmann-2012}, the
experimental data at $\theta=11.0^{\circ}$ shows
$P(b1)-P(b2)=0.857$ and $\Delta(P(b1)-P(b2))=0.00537$, which
corresponds to the minimum fluctuation in all experimental data 
obtained by changing $\theta$ discretely.
In comparison with $\theta = 2.2^{\circ}$, for example, 
they obtained $P(b1)-P(b2)=0.311$ and $\Delta(P(b1)-P(b2))=0.0131$.
The error at $\theta=2.2^{\circ}$ is 2.44 times as large as the error   
at $\theta=11.0^{\circ}$. This factor corresponds to 6 times
lower statistics at $\theta=11.0^{\circ}$, but the statistics at
any $\theta$ is almost same and the difference is absolutely
less than the factor two.
Taking into account the presence of noise sources, such as
imperfection of visibility, shot noise, etc., the minimum in
fluctuation at $\theta=11.0^{\circ}$ is consistent with the
above calculation.
However, no systematic experiments have been carried out yet for
the test of these theoretical predictions on fluctuations in
detail.


Finally, we give a comment on the other possibility of
interpretations of the ISTKH experiment.
In this paper, we choose the maximal entangled state as the
pre-selected state and the new problem arisen from the
indistinguishability between the system and the measuring device
was overcome by introducing the extended weak values.
Apart from such way, it may be also possible to assign the initial
polarization state (\ref{eq:init-state}) to the pre-selected
state in the same way as the ISTKH setup.
In this case, although it is not necessary to introduce the
extended weak values, we have to include the entanglement
creation into the four processes of the weak measurements.
Since the creation of the entangled state is also represented by
the unitary operator, the effective Hamiltonian corresponding to  
the unitary transformation included the creation process 
should be expressed as the specific mathematical form.
If we find this form, it will become possible to make the other 
re-interpretation along the original AAV argument with treating
the initial polarization state (\ref{eq:init-state}) as the
pre-selected state.
Although we leave this possibility of the interpretation as
future works, we note that many other interpretations for this
experiment will also be possible.


\section*{Acknowledgment}


We would like to thanks to Atsushi Nishizawa for his comments at
the beginning of this project. 
K.N. also thanks to Masaki Ando, Masa-Katsu Fujimoto, Akio
Hosoya for their comments and encouragement.




\begin{thebibliography}{99}
\bibitem{notation-comments}
  Although Stokes parameters are used to explain the photon
  polarization in ISTKH, throughout this paper, we describe the
  polarization states by the basis $\{|H\rangle,|V\rangle\}$ and
  we also introduce the operators
  $\hat{\sigma}_{x}=|H\rangle\langle V|+|V\rangle\langle H|$,
  $\hat{\sigma}_{y}= - i|H\rangle\langle V| + i|V\rangle\langle
  H|$, and $\hat{\sigma}_{z}= |H\rangle\langle H| -
  |V\rangle\langle V|$.
\bibitem{signature-comments}
  The difference of the overall sign in $P(b1)-P(b2)$ does not matter, 
  because the overall sign is experimentally controllable in the
  actual ISTKH experiment.
\end{thebibliography}
\end{document}